\documentclass[twocolumn, showpacs,preprintnumbers,amsmath,amssymb,prb,superscriptaddress]{revtex4-1}

\usepackage{graphicx}% Include figure files
\usepackage{dcolumn}% Align table columns on decimal point
\usepackage{bm}% bold math
\usepackage{epstopdf}
\usepackage{float}
\DeclareGraphicsExtensions{.pdf,.eps,.png,.jpg,.mps} 

\usepackage{color, soul}

\begin{document}

\title{Nuclear spin decoherence of neutral $^{31}$P donors in silicon: Effect of environmental $^{29}$Si nuclei}

\author{Evan S. Petersen}
 \affiliation{Department of Electrical Engineering, Princeton University}
\author{A. M. Tyryshkin}
 \affiliation{Department of Electrical Engineering, Princeton University}

\author{J. J. L. Morton}
\affiliation{London Centre for Nanotechnology, University College London}

\author{E. Abe}
\author{S. Tojo}
\author{K. M. Itoh}
\affiliation{Department of Applied Physics and Physico-Informatics, Keio University}

\author{M. L. W. Thewalt}
\affiliation{Department of Physics, Simon Fraser University}

\author{S. A. Lyon}
 \affiliation{Department of Electrical Engineering, Princeton University}

\begin{abstract}

Spectral diffusion arising from $^{29}$Si nuclear spin flip-flops, known to be a primary source of electron spin decoherence in silicon, is also predicted to limit the coherence times of neutral donor nuclear spins in silicon. Here, the impact of this mechanism on $^{31}$P nuclear spin coherence is measured as a function of $^{29}$Si concentration using X-band pulsed electron nuclear double resonance (ENDOR). The $^{31}$P nuclear spin echo decays show that decoherence is controlled by $^{29}$Si flip-flops resulting in both fast (exponential) and slow (non-exponential) spectral diffusion processes. The decay times span a range from 100~ms in crystals containing 50\% $^{29}$Si to 3~s in crystals containing 1\% $^{29}$Si. These nuclear spin echo decay times for \textit{neutral} donors are orders of magnitude longer than those reported for \textit{ionized} donors in natural silicon. The electron spin of the neutral donors `protects' the donor nuclear spins by suppressing $^{29}$Si flip-flops within a `frozen core', as a result of the detuning of the $^{29}$Si spins caused by their hyperfine coupling to the electron spin.

\end{abstract}

\maketitle

Donors in silicon have been considered for use in quantum information since the early days of the field.\cite{Kane1998, Vrijen2000} Donors have both electron and nuclear spins which can be manipulated independently, and both have been considered for use as potential quantum bits (qubits). While donor electron spins have received a majority of the attention,\cite{Morton2011,Tyryshkin2011,Pla2012,Zwanenburg2013} the nuclear spins are capable of much longer coherence times.\cite{Morton2008, Steger2012,Saeedi2013} This characteristic was utilized in the original Kane proposal for quantum computing\cite{Kane1998} and gained attention later for building a quantum memory.\cite{Morton2008} While exceptionally long T$_2$ times of donor nuclear spins in silicon have already been demonstrated,\cite{Morton2008,Steger2012} the mechanics of nuclear spin decoherence are not yet fully understood. In this study, we focus on neutral $^{31}$P donor nuclear spin decoherence arising from interactions with  $^{29}$Si nuclear spins in the silicon host environment.

Spectral diffusion due to spin 1/2 $^{29}$Si nuclei is a major source of decoherence for donor electron spins in silicon\cite{deSousa2003, Witzel2006, Abe2010,George2010} and is predicted to be a major source of decoherence for donor nuclear spins as well.\cite{deSousa2003} While the predicted coherence time for neutral $^{31}$P donor nuclear spins in natural silicon (containing 4.7\% $^{29}$Si) was 0.5~s, several experimental works reported much shorter times (from hundreds of microseconds to tens of milliseconds).\cite{Brown1974, Jeong2009,Dreher2012,Pla2013} Coherence times presented here and by Wolfowicz {\it et al.}\cite{Wolfowicz2015} show that the limit from $^{29}$Si spectral diffusion is actually longer than inferred from those previous experiments. To resolve the role of $^{29}$Si spectral diffusion, we measure neutral $^{31}$P nuclear spin coherence times in silicon crystals with $^{29}$Si concentrations ranging from 1\% to 50\%.\cite{Abe2010}

We find an inverse linear dependence of $^{31}$P nuclear spin coherence time on $^{29}$Si concentration ({\it f}), ranging from 100~ms at 50\% $^{29}$Si to 3~s at 1\% $^{29}$Si. The nuclear spin coherence time is about 1 second in natural silicon at 1.7~K; close to predictions of central spin stochastic models.\cite{deSousa2003} However, contrary to the predictions, the observed spin echo decays are non-exponential. The decay times are two orders of magnitude longer than those measured for ionized donors in natural silicon\cite{Dreher2012, Pla2013}  or in NMR experiments on degenerately doped silicon.\cite{Brown1974, Jeong2009} Apparently, the electron bound to a neutral donor protects the nuclear spin coherence from $^{29}$Si flip-flops by detuning nearby $^{29}$Si nuclear spins (a `frozen core').\cite{Khutsishvili1967, Wald1992, Guichard2015} This protection might not be required in high purity isotopically enriched silicon, with a low content of $^{29}$Si, where $^{29}$Si-induced spectral diffusion is no longer a dominant source of decoherence.\cite{Steger2012, Saeedi2013, Muhonen2014}

Four phosphorus-doped silicon crystals with different concentrations of $^{29}$Si isotopes were used in this work (Table~\ref{tab:table1}). In all crystals the donor concentration was about $10^{15}/{\rm cm}^3$ which is low enough to ensure that other decoherence effects arising from dipolar interactions with donor electron spins are small compared to the measured $^{29}$Si spectral diffusion effects. The pulsed ENDOR experiments were conducted using a Bruker Elexsys E580 spectrometer. Nuclear spin coherence times were measured using an electron-mediated nuclear spin Hahn echo experiment.\cite{Morton2008}  The combination of microwave and rf pulses enable a superposition state to be created on the donor electron, transferred to the $^{31}$P nucleus, manipulated on the nucleus, and then transferred back to the electron for readout. For temperatures below 5~K, when the electron T$_1$ relaxation was longer than 10~s, a light emitting diode (LED, 1050 nm) was flashed for 20~ms after each pulsed experiment in order to accelerate electron spin thermalization between repeated measurements. The ``tidy'' rf pulse to achieve nuclear spin thermalization was not required in these nuclear T$_2$ experiments.\cite{Tyryshkin2006, Morton2008} Most of the data shown were measured with a static magnetic field ($\sim$~0.35~T) oriented along a [001] crystal axis. Other field orientations were also examined to test the orientation dependence of the nuclear spin coherence times. 

\begin{table}%[H]
\begin{ruledtabular}
\begin{tabular}{lcccr}
Sample & $^{28}$Si (\%) & $^{29}$Si (\%) & $^{30}$Si (\%) & $^{31}$P/cm$^3$\\
\hline
\\
$^{29}$Si-1\% & 98.1 & 1.2 & 0.7 & $0.67\times10^{15}$\\
$^{29}$Si-5\% & 92.2 & 4.7 & 3.1 & $0.8\times10^{15}$\\
$^{29}$Si-10\% & 87.2 & 10.3 & 2.5 & $2.9\times10^{15}$\\
$^{29}$Si-50\% & 50.2 & 47.9 & 1.9 & $1.2\times10^{15}$\\
\end{tabular}
\end{ruledtabular}
\caption{\label{tab:table1} Four $^{31}$P-doped silicon samples used in this work. In each sample the $^{29}$Si concentration was determined by secondary ion mass spectroscopy (SIMS), and donor concentration was determined from ESR spin counting and independently confirmed by instantaneous diffusion slope measurements.\cite{Abe2010} All crystals were float-zone with the exception of $^{29}$Si-5\% (natural Si) which was Czochralski. All crystals had a volume on the order of a few cubic millimeters.  }
\end{table}

\begin{figure}
\centering
\includegraphics[scale=.6,keepaspectratio]{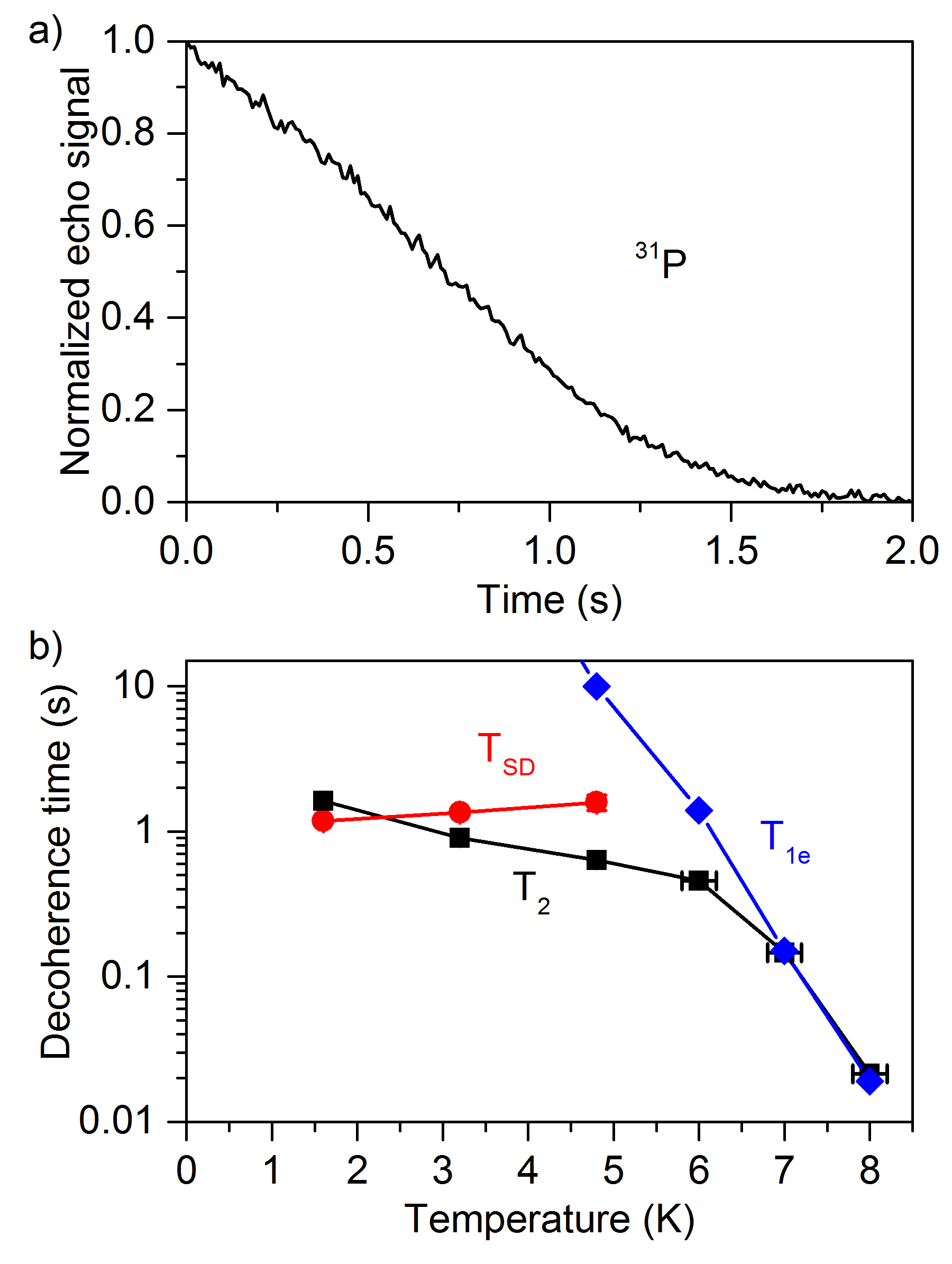}
\caption{\label{fig:decays} (color online) (a) Nuclear spin Hahn echo decays for neutral $^{31}$P donors in natural silicon ($f=4.7\%$) at 1.7~K with magnetic field ($\sim$~0.35~T) oriented along [001].  (b) Temperature dependences of $^{31}$P nuclear spin T$_2$ and T$_{\textrm{SD}}$ (black squares and red circles, respectively) and electron spin T$_{1e}$ times (blue diamonds) for neutral phosphorus donors in natural silicon.}
\end{figure}

The Hahn echo decay for phosphorus donor nuclear spins in natural silicon ($f=4.7\%$) at 1.7~K is shown in Fig. \ref{fig:decays}(a). This decay is non-exponential and can be best fit using:\cite{Zhidomirov1969,Chiba1972}
\begin{eqnarray}
v(\tau)=\exp\left(-\frac{2\tau}{T_2}-\left(\frac{2\tau}{T_{\textrm{SD}}}\right)^n\right)
\label{eq}
\end{eqnarray}
where $\tau$ is the time interval between $\pi$/2 and $\pi$ pulses in a Hahn echo experiment. This functional form contains two decoherence terms. T$_2$ can be associated with various decoherence processes, including T$_1$-related processes and a broad variety of spectral diffusion mechanisms in a {\it fast-motional} regime, while T$_{\textrm{SD}}$ is associated with spectral diffusion processes in a {\it slow-motional} regime.\cite{deSousa2003, Anderson1962, Mims1968} As we will discuss, fast and slow motional regimes in our experiments are defined by how the rates of $^{29}$Si nuclear spin flip-flops compare to the overall rate of decoherence. The stretch parameter $n$ is in the range between 2 and 3.\cite{Zhidomirov1969,deSousa2003,Chiba1972,Mims1968}

The temperature dependence of the extracted nuclear spin T$_2$ and T$_{\textrm{SD}}$ for phosphorous donors in natural silicon  is shown in Fig.~\ref{fig:decays}(b). The T$_{\textrm{SD}}$ term could only be extracted below 5~K because the linear T$_2$ term dominated the decays at higher temperatures. As seen from Fig.~\ref{fig:decays}(b), electron T$_1$ controls the nuclear T$_2$ at temperatures higher than 6~K, following the expected limit of T$_\textrm{2n}\sim 2\textrm{T}_{\textrm{1e}}$ within experimental errors.\cite{Morton2008} However, electron T$_1$ continues growing below that temperature, while both T$_2$ and T$_{\textrm{SD}}$ saturate at around 1~s showing little temperature dependence down to 1.7~K. This weak temperature dependence is consistent with $^{29}$Si-induced spectral diffusion being a dominant decoherence process for $^{31}$P nuclear spin below 5~K.\cite{deSousa2003} 

\begin{figure}
\centering
\includegraphics[scale=.38,keepaspectratio]{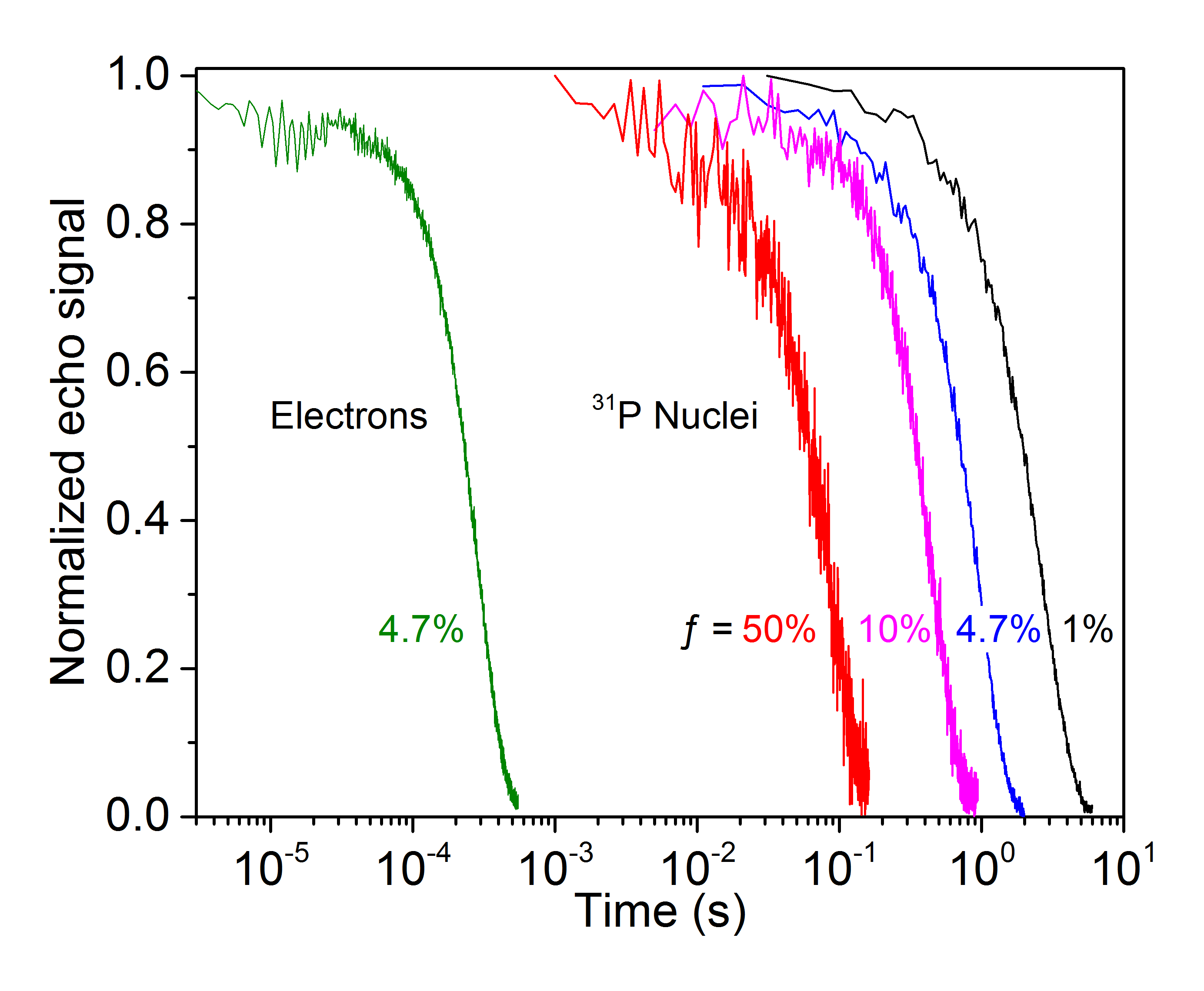}
\caption{\label{fig:decays2} (color online) $^{31}$P nuclear spin Hahn echo decays for phosphorus donors in silicon with different $^{29}$Si concentrations, measured at 1.7~K and magnetic field along [001]. The $^{29}$Si concentrations ($f$) are indicated for each curve. Electron spin Hahn echo decay for phosphorus donors in natural silicon is shown for comparison.}
\end{figure}

\begin{figure}
\centering
\includegraphics[scale=.6,keepaspectratio]{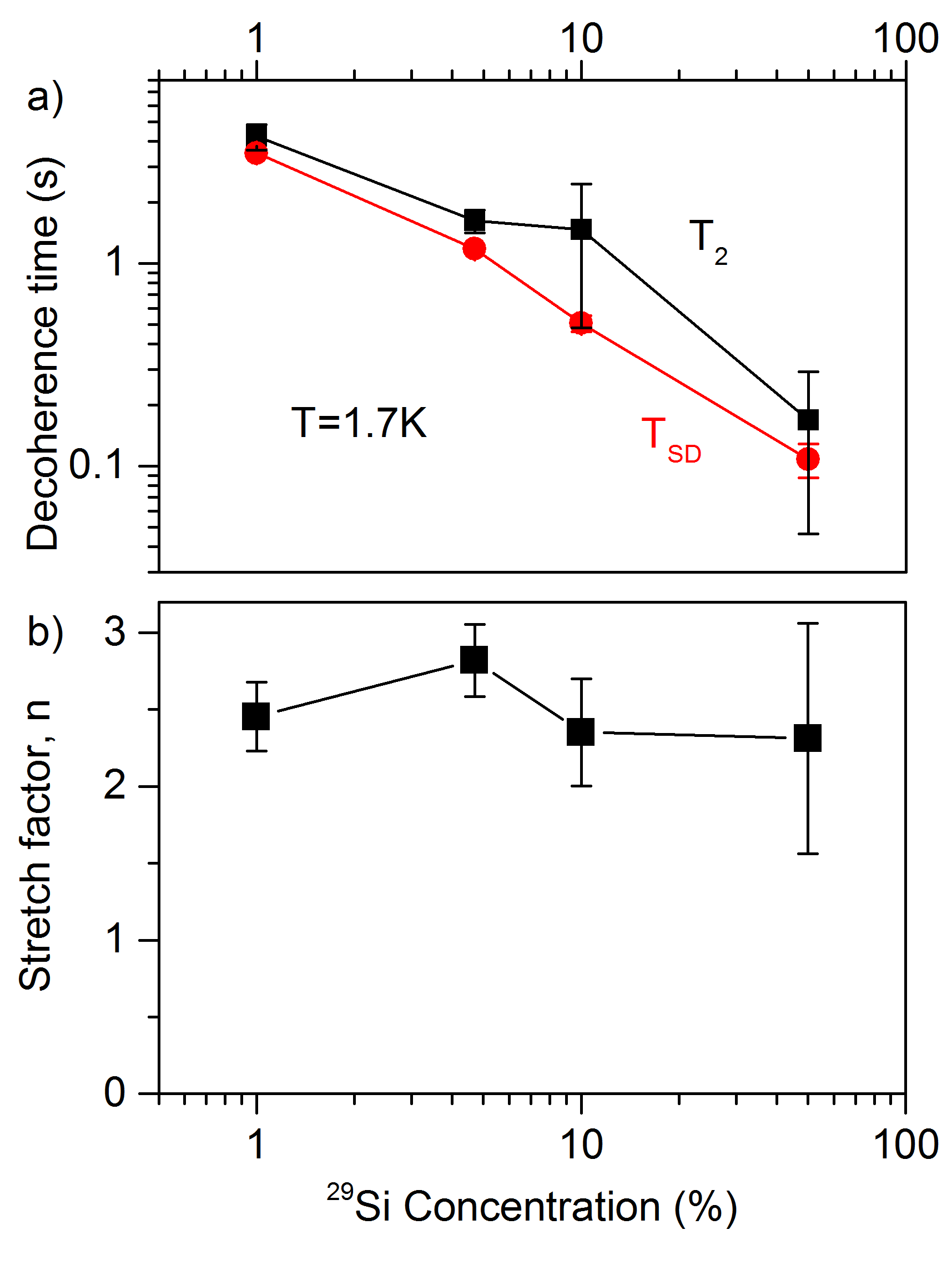}
\caption{\label{fig:params}(color online) $^{29}$Si concentration dependence of (a) spectral diffusion times T$_2$ and T$_{\textrm{SD}}$, and (b) spectral diffusion exponential power, $n$, for $^{31}$P nuclear spins of phosphorus donors in silicon at 1.7 K. Some error bars in (a) are smaller than their symbols.}
\end{figure}

The dependence of T$_2$ and T$_{\textrm{SD}}$ on $^{29}$Si concentration provides further evidence that $^{29}$Si flip-flops are a major source of decoherence in our samples. Nuclear spin echo decays for all four samples from Table~\ref{tab:table1} are shown in Fig.~\ref{fig:decays2} (electron spin echo decay for $^{31}$P donors in natural silicon is also shown for comparison). The extracted T$_2$ and T$_{\textrm{SD}}$ at 1.7~K are plotted against $^{29}$Si concentration in Fig.~\ref{fig:params}(a) showing a relatively inverse linear dependence for both times. Within the experimental errors, the parameter $n$, shown in Fig.~\ref{fig:params}(b), stays constant at about 2.5 for all concentrations from 1\% to 50\%. T$_2$ and T$_{\textrm{SD}}$ were also measured with the magnetic field oriented at different angles with respect to the crystal axis. However, no noticeable orientation dependence was observed within experimental errors (10\%) in either T$_2$ or T$_{\textrm{SD}}$. 

Three other decoherence mechanisms must be considered here as potential contributors to nuclear spin decoherence at low temperatures. These processes have been found to be significant in decohering electron spins of neutral donors. All three mechanisms are related to dipolar interactions with electron spins of other donors. The first two processes are cases of spectral diffusion arising (1) from T$_1$-driven flips of electron spins of nearby donors,\cite{Mims1968,Tyryshkin2011} and (2) from electron spin flip-flops in nearby donor pairs.\cite{Witzel2010,Tyryshkin2011}  Both cases are much less effective in decohering nuclear spins than electron spins since their effect scales proportionally with nuclear and electron gyromagnetic ratios ($\sim$1/1,600 in the case of $^{31}$P nuclei).  Using the electron T$_2$ times reported in Ref.~[\onlinecite{Tyryshkin2011}], assuming donor densities of $10^{15}/{\rm cm}^3$ (as used here) and considering temperatures below 4.8~K, we can then estimate the nuclear T$_2$ and T$_{\textrm{SD}}$ from spectral diffusion processes (1) and (2) to be longer than 200~s. Thus, processes (1) and (2) are too slow to explain our $T_2$ data in Fig. ~\ref{fig:params}(a).

The third dipolar-related process to be considered is a ``direct" flip-flop process.\cite{Kurshev1992} This involves a spin flip-flop between an electron of a ``central" donor and an electron of a neighboring donor (this is in contrast to ``indirect" flip-flops described in (2) above). Unlike other dipolar-related mechanisms, the effect of direct flip-flops does not scale with gyromagnectic ratio, decohering nuclear spins as fast as electron spins. Direct flip-flops have been reported to limit electron spin coherence to 0.6 s for donors at $10^{14}/{\rm cm}^3$ in isotopically-purified $^{28}$Si crystals (45 ppm of $^{29}$Si).\cite{Tyryshkin2011} However, the inhomogeneous broadening in our samples (Table~\ref{tab:table1}) is 100-900~$\mu$T which is 20-200 times broader than the 5~$\mu$T found in the aforementioned 45~ppm crystals.\cite{Abe2010,Tyryshkin2011} Taking into account the donor density in our samples we estimate that the direct flip-flop contribution to $^{31}$P nuclear decoherence is about 3~s in our $^{29}$Si-1\% sample and is even longer ($>$10~s) in the other three samples.

The effect of $^{29}$Si spectral diffusion on nuclear spin coherence of neutral $^{31}$P donors in silicon has been examined theoretically in the framework of a central spin problem while modeling $^{29}$Si spin flip-flops as a classical stochastic process.\cite{deSousa2003}  For natural silicon a $^{31}$P nuclear spin coherence time of 0.5~s  was predicted with the field oriented along [001]. This prediction is very close to what was measured at that orientation in this work. Simulations of nuclear spin coherence showed an approximately inverse dependence on $f$, which also correlates with our results. However, the theory predicted exponential T$_2$ decays with $n=1$, while our experiment shows non-exponential decays. The predicted orientation dependence was also not observed in our experiment.

It is instructive to compare the decoherence of $^{31}$P donor electron and nuclear spins caused by $^{29}$Si flip-flops. The electron and nuclear spin echo decays differ in two ways: (i) the nuclear spin echo decays are over 3 orders of magnitude longer than electron spin echo decays, and (ii) the nuclear spin echo decays contain both exponential and non-exponential components, unlike the electron spin echo decays that are dominated by only the non-exponential term.\cite{Abe2010} Both electrons and $^{31}$P nuclei see the same bath of $^{29}$Si nuclear spins, with the bath's dynamics suppressed in the ``frozen core'' where the nuclear spins are detuned by the donor electron.\cite{Khutsishvili1967, Wald1992, Guichard2015} The main difference between electron and $^{31}$P nuclear spins is the strength of their interactions to the $^{29}$Si spin bath. Contact hyperfine interactions for an electron spin are much stronger than dipolar interactions for a $^{31}$P nuclear spin, therefore the same $^{29}$Si bath decoheres the electron spin faster than the nuclear spin. This difference in the coherence timescale explains (i) and is also the key to understanding (ii).

$^{29}$Si flip-flops can cause fast or slow motional effects depending on whether the flip-flop rate is fast or slow compared to the strength of the pair's interaction with the central spin.\cite{deSousa2003, Anderson1953, Zhidomirov1969, Mims1968} Equivalently, the fast and slow regimes can be descriminated by comparing the flip-flop period to the overall coherence timescale (2$\tau$) of the central spin.\cite{Mims1968} There is a broad distribution of $^{29}$Si flip-flop rates, with the fastest rates being $\sim$~100~Hz and 10~Hz in nearest and next-nearest neighbor pairs, and much slower rates in more distant pairs. All these rates correspond to times much longer than 2$\tau$ ($\sim 600~\mu s$) when measuring electron spin echoes. In this case all flip-flops are in a slow motional regime, causing slow spectral diffusion with non-exponential echo decays as seen in experiment\cite{Abe2010} and understood theoretically.\cite{Chiba1972, deSousa2003, Witzel2012} For nuclear spins, on the other hand, the timescale of the experiments lies within the broad distribution of $^{29}$Si flip-flop times. Thus, there are fast and slow flip-flopping pairs that contribute to the decoherence, and consistently both exponential and non-exponential components are present in the decays.

The $^{31}$P nuclear spin coherence time in natural silicon presented here is longer than measured earlier for ionized donors or donors in degenerately doped silicon. NMR measurements of $^{31}$P nuclear spin decoherence in degenerately doped silicon have found times about two orders of magnitude shorter.\cite{Brown1974, Jeong2009} Ionized donors measured with EDMR had a coherence time of 18~ms,\cite{Dreher2012} and single donors measured with an SET had a coherence time of 60~ms.\cite{Pla2013} These measurements of ionized donors are in agreement with cluster correlation expansion simulations by Witzel {\it et al.} ($\sim$30~ms).\cite{Witzel2012} The longer coherence time for our isolated neutral $^{31}$P donors supports the ``frozen core" picture\cite{Khutsishvili1967, Wald1992, Guichard2015} where most $^{29}$Si pairs near a central spin are too detuned by the donor electron spin to flip-flop.

In conclusion, we have experimentally studied the effect of environmental $^{29}$Si nuclear spins on neutral donor nuclear spins decoherence in silicon. Two contributors have been resolved arising from fast and slow flip-flopping $^{29}$Si nuclear spin pairs. We find that both contributions exhibit a linear dependence on $^{29}$Si concentration. Our results demonstrate long coherence times for neutral donor nuclear spins, ranging from 100~ms in crystals containing 50\% $^{29}$Si to 3~s in crystals containing 1\% $^{29}$Si, and are in agreement with the picture that  an electron bound to a donor protects the donor nuclear spins from $^{29}$Si flip-flops. 

Work was supported by the NSF through the Materials World Network, EPSRC, and MRSEC Programs (Grant No. DMR-1107606, EP/I035536/1, and DMR-01420541), and the ARO (Grant No. W911NF-13-1-0179). The research at Keio was supported part by MEXT/JSPS KAKENHI No. 26220602, JSPS Core-to-Core Program,  and NanoQuine. J.J.L.M. is supported by the Royal Society.

\bibliography{espFull}
\end{document}